\documentclass[a4paper,10pt]{article}
\usepackage[dvips]{graphicx}
\usepackage{amssymb,amsmath}
\numberwithin{equation}{section}

\begin{document}

\title{  G-essence with  Yukawa Interactions}
\author{ I.Kulnazarov$^1$, K.Yerzhanov$^1$, O.Razina$^1$, Sh.Myrzakul$^1$, P.Tsyba$^1$,    R.Myrzakulov$^{1,2}$\footnote{The corresponding author. Email: rmyrzakulov@gmail.com; rmyrzakulov@csufresno.edu}\vspace{1cm}\\ \textit{$^1$Eurasian International Center for Theoretical Physics,} \\ \textit{Eurasian National University, Astana 010008, Kazakhstan}\\ \textit{$^2$Department of Physics, CSU Fresno, Fresno, CA 93740 USA}}

\date{}

\maketitle
\begin{abstract} We study the g-essence model with Yukawa interactions between  a scalar field $\phi$ and a Dirac field $\psi$. For the  homogeneous, isotropic and flat Friedmann--Robertson--Walker universe 
 filled with the such g-essence, the  exact solution of the model is found. Moreover, we reconstruct the corresponding scalar and fermionic potentials which describe the coupled dynamics of the scalar and fermionic   fields. It is  shown that some particular g-essence models with Yukawa  interactions correspond to the usual and generalized Chaplygin gas unified models of dark energy and dark matter.  Also we present some scalar--fermionic Dirac--Born--Infeld  models corresponding g-essence models with Yukawa interactions which again describe the unified dark energy--dark matter system.
\end{abstract}
\vspace{2cm} 

\sloppy

\section{Introduction} One of the most puzzling discovery of the last years in physics is the current acceleration of the universe \cite{Perlmutter}--\cite{Riess}. An unknown energy component, dubbed as dark energy, is proposed to explain this acceleration.  Dark energy almost equally distributes in the universe, and its pressure is negative. The simplest and most theoretically appealing candidate of dark energy is the cosmological constant that is the $\Lambda$CDM model. In this case, the equation of state parameter $\omega=-1$. Although the $\Lambda$CDM model is in general agreement with the current astronomical observations, but has some difficulties e.g. to reconcile the small observational value of dark energy density with estimates from quantum field theories. So although  the $\Lambda$CDM model is the most obvious choice, but it suffers from coincidence problem and the fine-tuning problems.  It is thus natural to pursue alternative possibilities to explain the mystery of dark energy. In order to explain the acceleration that is dark energy, many kinds of models have been proposed, such as quintessence, phantom, k-essence, tachyon, f-essence, Chaplygin gas and its generalizations, etc.

 In the last years, the k-essence model has received much attention. It was originally proposed as a model for inflation \cite{Mukhanov1}, and then as a model for dark energy \cite{Mukhanov2}--\cite{Linder}.  Since from it was proposed, k-essence was been studied intensively. It is still worth investigating in a systematic way the possible cosmological behavior of the k-essence.  Quite recently, the so-called g-essence model has been proposed  \cite{MR}, which is a more generalized model than k-essence. In fact, the g-essence contains,  as particular cases, two important models: k-essence and f-essence. Note that  f-essence is the fermionic counterpart of k-essence. 
 
 To our knowledge, in the literature there are relatively few works on the dark energy models with fermionic fields. However, in the recent years several approaches were made to explain the accelerated expansion by choosing fermionic fields as the gravitational sources of energy (see e.g. refs. \cite{Ribas1}--\cite{Armendariz-Picon}). In particular, it was shown that the fermionic  field plays very important role in: i) isotropization of initially anisotropic spacetime; ii) formation of singularity free cosmological solutions; iii) explaining late-time acceleration. In particle physics, the  Yukawa interaction  plays an important role. It has the form
 \begin {equation}
U=-g\bar{\psi}\phi\psi.
\end{equation} 

 It   describes the interaction  between  a scalar field $\phi$ and a Dirac field $\psi$. Some properties of the Yukawa interaction (1.1) related with the gravitational field were considered in \cite{Zanusso}--\cite{Woodard2}. With the Yukawa interaction (1.1) is related the so-called  Yukawa potential,
  \begin {equation}
V(x)=-g^2x^{-1}e^{-mx}
\end{equation} 
which is negative, that is, the corresponding force is attractive. The relation between the Yukawa potential (1.2) and the accelerated expansion of the universe were studied by some authors (see e.g. \cite{Atazadeh}). In this paper, we focus on so-called g-essence model \cite{MR} which is some hybrid construction of k-essence and f-essence. If exactly,  we will consider the g-essence with the Yukawa interaction (1.1). The  formulation of the gravity-fermionic theory has been discussed in detail elsewhere \cite{Weinberg}--\cite{Birrell}., so we will only present the result here.

 This paper is organized as follows. In the following section, we briefly review g-essence. In Sec. 3, we introduce the g-essence model with the Yukawa interaction. In Sec. 4, we construct the solution of the particular g-essence model with the scalar--fermionic Yukawa interaction. The unified scalar--fermionic Chaplygin gas model of dark energy and dark matter from the g-essence model with the Yukawa interaction were constructed in section 5 and its extension for the generalized Chaplygin gas case in section 6. The scalar--fermionic Dirac--Born--Infeld (DBI) counterpart of the g-essence model with the Yukawa interaction (3.1) was constructed in section 7.   Finally, we shall close with a few concluding remarks in Sec. 8. The metric  signature used is $(+, -, -, -)$ and units have been chosen so that $8\pi G=c=\hbar=1.$
\section{Basics of g-essence}

The action of   g-essence has the form \cite{MR}
\begin {equation}
S=\int d^{4}x\sqrt{-g}[R+2K(X, Y, \phi, \psi, \bar{\psi})],
\end{equation} 
 where $K$ is the  g-essence  Lagrangian and is some function of its arguments, $\phi$ is a scalar function, $\psi=(\psi_1, \psi_2, \psi_3, \psi_4)^{T}$  is a fermionic function  and $\bar{\psi}=\psi^+\gamma^0$ is its adjoint function, the curvature scalar $R$. Here
\begin {equation}
X=0.5g^{\mu\nu}\nabla_{\mu}\phi\nabla_{\nu}\phi,\quad Y=0.5i[\bar{\psi}\Gamma^{\mu}D_{\mu}\psi-(D_{\mu}\bar{\psi})\Gamma^{\mu}\psi]
\end{equation}
are  the canonical kinetic terms for the scalar and fermionic fields, respectively. $\nabla_{\mu}$ and  $ D_{\mu}$ are the covariant derivatives. The fermionic fields are treated here as classically commuting fields.  The model (2.1) admits important two reductions: \textit{k-essence} and \textit{f-essence}. In this sense,  it is the more general essence model and in \cite{MR} it was   called \textit{g-essence}. 
Note that to find the equations of motion we need the variations
\begin{eqnarray}
	\delta\sqrt{-g}&=&-0.5g_{\mu\nu}\sqrt{-g}\delta g^{\mu\nu},\\ 
		\delta R&=&(R_{\mu\nu}+g_{\mu\nu}\square-\nabla_{\mu}\nabla_{\nu})\delta g^{\mu\nu},\\
		\delta K&=&K^{'}\delta\phi+K^{''}\delta\psi+K^{'''}\delta\bar{\psi},
	\end{eqnarray} 
	where $\nabla_{\nu}V_{\mu}\equiv\partial_{\nu}V_{\mu}-\Gamma^{\sigma}_{\mu\nu}V_{\sigma}$ is the covariant derivative of a vector $V_{\mu}$ and the curved d'Alembertian on a scalar $\phi$ is
		\begin{equation}
\square\phi=\frac{1}{\sqrt{-g}}\partial^{\mu}(\sqrt{-g}\partial_{\mu}\phi).
\end{equation}
We now  consider the dynamics of the  homogeneous, isotropic and flat FRW universe filled with g-essence. In this case, the background line element reads
	\begin{equation}
ds^2=dt^2-a^2(dx^2+dy^2+dz^2)
\end{equation}
and the vierbein is chosen to be
	\begin{equation}
	(e_a^\mu)=diag(1,1/a,1/a,1/a),\quad 
(e^a_\mu)=diag(1,a,a,a).
\end{equation}

 In the case of the FRW metric (2.7), the equations corresponding to the action (2.1) look  like \cite{MR}
	\begin{eqnarray}
	3H^2-\rho&=&0,\\ 
		2\dot{H}+3H^2+p&=&0,\\
		K_{X}\ddot{\phi}+(\dot{K}_{X}+3HK_{X})\dot{\phi}-K_{\phi}&=&0,\\
		K_{Y}\dot{\psi}+0.5(3HK_{Y}+\dot{K}_{Y})\psi-i\gamma^0K_{\bar{\psi}}&=&0,\\ 
K_{Y}\dot{\bar{\psi}}+0.5(3HK_{Y}+\dot{K}_{Y})\bar{\psi}+iK_{\psi}\gamma^{0}&=&0,\\
	\dot{\rho}+3H(\rho+p)&=&0,
	\end{eqnarray} 
where  $K_X=dK/dX, \quad K_Y=dK/dY,\quad K_{\phi}=dK/d\phi,\quad K_{\psi}=dK/d\psi, \quad K_{\bar{\psi}}=dK/d\bar{\psi}$ and  $H=\dot{a}/a$ denotes the Hubble parameter, the dot represents a differentiation with respect to time $t$. Here the kinetic terms, the energy density  and  the pressure  take the form
\begin{equation}
X=0.5\dot{\phi}^2,\quad  Y=0.5i(\bar{\psi}\gamma^{0}\dot{\psi}-\dot{\bar{\psi}}\gamma^{0}\psi)
  \end{equation}
  and 
\begin{equation}
\rho=2XK_{X}+YK_{Y}-K,\quad
p=K.
\end{equation}

 Note that the equations of g-essence (2.9)-(2.14) can be rewritten as	
	\begin{eqnarray}
	3H^2-\rho&=&0,\\ 
		2\dot{H}+3H^2+p&=&0,\\
		(a^3K_{X}\dot{\phi})_{t}-a^3K_{\phi}&=&0,\\
		(a^3K_{Y}\psi^2_j)_{t}-2iK_{\bar{\psi}}(\gamma^0\psi)_j&=&0,\\ 
(a^3K_{Y}\psi^{*2}_j)_{t}+2iK_{\psi}(\bar{\psi}\gamma^{0})_j&=&0,\\
	\dot{\rho}+3H(\rho+p)&=&0.
	\end{eqnarray} 
	Also  we present the useful formula
	\begin{equation}
K_{Y}Y=0.5iK_{Y}(\bar{\psi}\gamma^{0}\dot{\psi}-\dot{\bar{\psi}}\gamma^{0}\psi)=-0.5(K_{\psi}\psi+K_{\bar{\psi}}\bar{\psi})
  \end{equation}
  and the equation for $u=\bar{\psi}\psi$:
  \begin{equation}
[\ln{(ua^3K_Y)}]_tu=-iK^{-1}_{Y}(\bar{\psi}\gamma^{0}K_{\bar{\psi}}-K_{\psi}\gamma^0\psi).
  \end{equation}
Finally,  we note that some exact solutions of g-essence (2.9)--(2.14) were presented in \cite{MR3}-\cite{MR30} (see, also \cite{MR71}-\cite{MR73} in which some integrable aspects of the FRW models were considered).

\subsection{Purely kinetic g-essence}
Let us consider the purely kinetic g-essence,  that is, when $K=K(X,Y)$. In this case, the system (2.9)--(2.14) becomes		
	\begin{eqnarray}
	3H^2-\rho&=&0,\\ 
		2\dot{H}+3H^2+p&=&0,\\
		a^3K_{X}\dot{\phi}-\sigma &=&0,\\
		a^3K_{Y}\psi^2_j-\varsigma_j &=&0,\\ 
a^3K_{Y}\psi^{*2}_j-\varsigma^{*}_j&=&0,\\
	\dot{\rho}+3H(\rho+p)&=&0,
	\end{eqnarray}
	where $\sigma$  $(\varsigma)$ is the real (complex) constant. Hence we immediately get the solutions of the  Klein--Gordon and Dirac equations, respectively, as
	\begin{equation}
\phi=\sigma\int\frac{dt}{ a^3K_{X}},\quad \psi_j=\sqrt{\frac{\varsigma_j}{a^3K_{Y}}}.
\end{equation}
 Also the following useful formula holds:
 	\begin{equation}
X=\frac{0.5\sigma^2}{a^6K_{X}^2} \quad or \quad K_X=\frac{\sigma}{a^3\sqrt{2X}}.
\end{equation}

\subsection{K-essence}

Let us now we consider the following particular case of g-essence
 (2.1):
\begin{equation}
K=K_1=K_1(X,\phi)
\end{equation} that corresponds to k-essence.
Then the system (2.9)--(2.14) takes the form of the equations of   k-essence (see e.g.  \cite{Mukhanov1}--\cite{Chiba})
\begin{eqnarray}
	3H^2-\rho_k &=&0,\\ 
		2\dot{H}+3H^2+p_k&=&0,\\
		K_{1X}\ddot{\phi}+(\dot{K}_{1X}+3HK_{1X})\dot{\phi}-K_{1\phi}&=&0,\\
	\dot{\rho}_k+3H(\rho_k+p_k)&=&0,
	\end{eqnarray} 
where  the energy density  and  the pressure  are given by
\begin{equation}
\rho_k=2XK_{1X}-K_1,\quad
p_k=K_1.
\end{equation} 
As is well known, the  energy-momentum tensor for the k-essence field has the form
	\begin{equation}
T_{\mu\nu}=K_{X}\nabla_{\mu}\phi\nabla_{\nu}\phi
	-g_{\mu\nu}K=2K_{X}Xu_{1\mu}u_{1\nu}-Kg_{\mu\nu}=(\rho_k+p_k)u_{1\mu}u_{1\nu}-p_kg_{\mu\nu}.
\end{equation}

It is interesting to note that in the case of  the FRW metric (2.7), purely kinetic k-essence and F(T)-gravity (modified teleparallel gravity) are equivalent  to each other, if $a=e^{\pm\frac{\phi-\phi_0}{\sqrt{12}}}$ \cite{MR1}--\cite{MR2}.

\subsection{F-essence}

Now we consider the following reduction of g-essence  (2.1):
\begin{equation}
K=K_2=K_2(Y,\psi, \bar{\psi})
\end{equation}
that is,  f-essence \cite{MR}. The  energy-momentum tensor for the f-essence field has the form
$$
T_{\mu\nu}\equiv\frac{2}{\sqrt{-g}}\frac{\delta S}{\delta g_{\mu\nu}}=0.5iK_{Y}\left[\bar{\psi}\Gamma_{(\mu}D_{\nu)}\psi-D_{(\mu}\bar{\psi}\Gamma_{\nu)}\psi\right]-
$$
	\begin{equation}	-g_{\mu\nu}K=K_{Y}Yu_{2\mu}u_{2\nu}-Kg_{\mu\nu}=(\rho_f+p_f)u_{2\mu}u_{2\nu}-p_fg_{\mu\nu}.
\end{equation}
For the FRW metric (2.7),  the equations of the f-essence  become \cite{MR}
\begin{eqnarray}
	3H^2-\rho_f &=&0,\\ 
		2\dot{H}+3H^2+p_f&=&0,\\
		K_{2Y}\dot{\psi}+0.5(3HK_{2Y}+\dot{K}_{2Y})\psi-i\gamma^0K_{2\bar{\psi}}&=&0,\\ 
K_{2Y}\dot{\bar{\psi}}+0.5(3HK_{2Y}+\dot{K}_{2Y})\bar{\psi}+iK_{2\psi}\gamma^{0}&=&0,\\
	\dot{\rho}_f+3H(\rho_f+p_f)&=&0,
	\end{eqnarray} 
where 
\begin{equation}
\rho_f=YK_{2Y}-K_2,\quad
p_f=K_2.
\end{equation}

Some properties of f-essence were studied in \cite{MR01}-\cite{MR02}. 
\section{Model} 
In this section, we consider   the  action (2.1) with the following particular g-essence Lagrangian:
\begin {equation}
K=\alpha_1 X+ \alpha_2 X^n+\alpha_3 V_{1}(\phi)+\beta_1 Y+\beta_2 Y^m+\beta_3 V_2(\bar{\psi}, \psi)+\eta U_1(\phi)U_2(\bar{\psi}, \psi),
\end{equation} 
where $\alpha_j, \beta_j, \eta$ are some real constants. As the search for exact solutions of the coupled system of  differential equations (2.9)--(2.14) for the g-essence Lagrangian (3.1) is a very hard job,  let us simplify the problem e.g. as $V_{2}=V_{2}(u), \beta_2=0, U_1=\phi, U_2=u	$.	Then the  system  (2.9)--(2.14) takes the form
		\begin{eqnarray}
	3H^2-\rho&=&0,\\ 
		2\dot{H}+3H^2+p&=&0,\\
		\ddot{\phi}+[3H+(\ln{(\alpha_1+\alpha_2n X^{n-1})})_{t}]\dot{\phi}-\frac{\alpha_3V_{1\phi}+\eta u}{\alpha_1+\alpha_2n X^{n-1}}&=&0,\\
		\dot{\psi}+1.5H\psi-i\beta^{-1}_{1}\gamma^0(\beta_3V_{2u}\psi+\eta \phi\psi)&=&0,\\ 
\dot{\bar{\psi}}+1.5H\bar{\psi}+i\beta^{-1}_1(\beta_3V_{2u}\bar{\psi}+\eta \phi\bar{\psi})\gamma^0&=&0,\\
	\dot{\rho}+3H(\rho+p)&=&0,
	\end{eqnarray} 
	where
	\begin{eqnarray}
\rho&=&\alpha_1X+\alpha_2(2n-1)X^n-\alpha_3V_1-\beta_3V_2-\eta\phi u, \\ p&=&\alpha_1 X+ \alpha_2 X^n+\alpha_3 V_{1}+\beta_1 Y+\beta_3 V_2+\eta \phi u.
\end{eqnarray}
	
Hence and from (2.23)--(2.24)  we get
	\begin{eqnarray}
\rho+p&=&2\alpha_1X+2\alpha_2nX^n+\beta_1Y=-2\dot{H}, \\ \beta_1Y&=&-(\beta_3V^{'}_{2}+\eta\phi)u,\\
 u&=&\frac{c}{\beta_1 a^3}.
\end{eqnarray}

\section{Solution}

In this section, we want to present the exact solution of the system (3.2)--(3.7). But first note that for six unknown functions $a, \phi, \psi, \bar{\psi}, V_1, V_2$ we have five differential equations (3.2)--(3.6) so that we need one more equation (see e.g. \cite{MR4}--\cite{Odintsov}). Such an equation we take to be
\begin{equation}
a=\bar{a}_0\phi^{k},
\end{equation}	
where $\bar{a}_{0}=a_{0}\phi_0^{-k}, \quad k=\lambda/\delta$. Then we obtain  the following solution
\begin{eqnarray}
a&=&a_0t^{\lambda},\\ 
\phi&=&\phi_0t^{\delta},\\ 
\psi_l&=&\frac{c_{l}}{a^{1.5}_0t^{1.5\lambda}}e^{-iD},\quad (l=1,2),\\	
\psi_k&=&\frac{c_{k}}{a^{1.5}_0t^{1.5\lambda}}e^{iD}, \quad (k=3,4),
	\end{eqnarray}
where  $c_j$ obey the condition
	\begin{equation}
c=|c_{1}|^2+|c_{2}|^2-|c_{3}|^2-|c_{4}|^2
\end{equation}
 and
 $$
D=\frac{2}{\beta_1u_0[2(\delta-1)+3\lambda+1]}t^{2(\delta-1)+3\lambda+1}+
$$
\begin{equation}
\frac{2^{1-n}\alpha_2n\delta^{2n}\phi_0^{2n}}{u_0[2n(\delta-1)+3\lambda+1]}t^{2n(\delta-1)+3\lambda+1}-\frac{2\lambda}{u_0(3\lambda-1)}t^{3\lambda-1}+D_0.
\end{equation}
This solution is correct if 
 	\begin{equation}
\delta=2-3\lambda
\end{equation}
or
 	\begin{equation}
\delta=\frac{3\lambda-2n}{1-2n}.
\end{equation}
The corresponding  potentials  take   the form
\begin{eqnarray}
V_1(\phi)&=&l_1\left(\frac{\phi}{\phi_0}\right)^{\frac{2(\delta-1)}{\delta}}+l_2\left(\frac{\phi}{\phi_0}\right)^{\frac{2n(\delta-1)}{\delta}}+l_3\left(\frac{\phi}{\phi_0}\right)^{\frac{\delta-3\lambda}{\delta}}+V_{10},\\ 
V_2(u)&=&q_1\left(\frac{u}{u_0}\right)^{\frac{2(1-\delta)}{3\lambda}}+q_2\left(\frac{u}{u_0}\right)^{\frac{2n(1-\delta)}{3\lambda}}+q_3\left(\frac{u}{u_0}\right)^{\frac{3\lambda-\delta}{3\lambda}}+q_4\left(\frac{u}{u_0}\right)^{\frac{2}{3\lambda}}-\alpha_3\beta_3^{-1}V_{10}.
	\end{eqnarray}
Here $V_{10}=const,\quad u_0=ca_0^{-3}\beta^{-1}_1$,
\begin{eqnarray}
l_1&=&\frac{\alpha_1\delta^2\phi_0^2(\delta-1+3\lambda)}{2\alpha_3(\delta-1)},\\ 
l_2&=&\frac{\alpha_2\delta^{2n}\phi_0^{2n}[(2n-1)(\delta-1)+3\lambda]}{2^n\alpha_3(\delta-1)},\\ 
l_3&=&-\frac{\eta\delta\phi_0u_0}{\alpha_3(\delta-3\lambda)}
	\end{eqnarray}
	and
	\begin{eqnarray}
q_1&=&0.5\alpha_1\delta^2\phi_0^2-\alpha_3l_1,\\ 
q_2&=&\alpha_2(2n-1)2^{-n}\delta^{2n}\phi_0^{2n}-\alpha_3l_2,\\ 
q_3&=&-\alpha_3l_3-\eta\phi_0u_0,\\
q_4&=&-3\lambda^2.
	\end{eqnarray}
	
	\section{Unified scalar--fermionic Chaplygin gas  model of dark energy and dark matter from g-essence with Yukawa interactions}The most popular models of dark energy   and dark matter  such as  e.g. the $\Lambda$CDM	and a quintesse-CDM model assume that dark energy   and dark matter  are distinct entities. Another interpretation of the observational data is that dark energy   and dark matter  are different manifestations of a common structure. The first definite model of this type was proposed in \cite{Kamenshchik}, based upon the Chaplygin gas, an exotic  perfect fluid obeying the equation of state (EoS)
		\begin{equation}
p=-\frac{A}{\rho},
\end{equation}
which has been extensively studied for its mathematical properties \cite{Jackiw}.	The general class of models, in which a unification of dark energy   and dark matter  is achieved through a single entity, is often referred to as quartessence. Among other scenarios of unification that have recently been suggested, interesting attempts are based on k-essence. In this section we  extend this scenario to  the   g-essence model with Yukawa interactions which gives us the Chaplygin gas unified model of dark energy   and dark matter. To do it, let us consider the  g-essence model  given by the following  scalar--fermionic DBI Lagrangian:
		\begin{equation}
K=U\sqrt{1+V_1 X+ V_2Y^2},
\end{equation} 
where in general $U=U(\phi,\bar{\psi},\psi), \quad V_1=V_1(\phi,\bar{\psi},\psi),\quad V_2=V_2(\phi,\bar{\psi},\psi).$ 
Note that the g-essence model (5.2) is constrained in two particular cases: i) the scalar DBI model as $U=U(\phi), \quad V_1=V_1(\phi), \quad V_2=0$; ii) the fermionic DBI model as $U=U(\bar{\psi},\psi),\quad V_1=0,\quad  V_2=V_2(\bar{\psi},\psi)$. Substituting the expression (5.2) into (2.16)  we get 
	\begin{eqnarray}
p&=&U\sqrt{1+V_1 X+ V_2Y^2},\\ 
\rho&=&-\frac{U}{\sqrt{1+V_1 X+ V_2Y^2}},	\end{eqnarray}
where  	we assume that	
\begin{equation}
U=\eta \phi u, \quad V_1=V_1(\phi,\bar{\psi},\psi),\quad V_2=V_2(\phi,\bar{\psi},\psi).
\end{equation}
It is  the   g-essence model with Yukawa interactions $U=\eta \phi u$. These equations give the following EoS:
  	\begin{equation}
p=-\frac{ U^2}{\rho}.
\end{equation}

  It is the Chaplygin gas model \cite{Kamenshchik} but with the variable function $U$ (Yukawa interactions). From (5.6) and (2.14), we get
  	\begin{equation}
\rho=a^{-3}\left[6\int U^2(a)a^5da+C\right]^{0.5}=z^{-0.5}\left[C+\int U^2(z)dz\right]^{0.5}, 
\end{equation} 
where $C=const, \quad z=a^6$.	From these formulas we obtain the following expression for the EoS parameter:
	\begin{equation}
\omega=-\frac{ U^2}{\rho^2}=-1-\frac{zd\ln \rho^2}{dz}=-1-\frac{d\ln \rho^2}{d\ln z}=-\frac{ zU^2}{C+\int U^2(z)dz}=-\frac{ zh^{'}}{C+h}=-z[\ln(C+h)]^{'},
\end{equation}
where $h=	\int U^2(z)dz,\quad h^{'}=dh/dz$. In principle, now it is not difficult to construct  solutions of the g-essence   equations corresponding to the different expressions for $U.$ Here we just present the expressions for the energy density and pressure.  Consider some examples.

i) Let $U=\mu=const$. Then from (5.7) and (5.6) we obtain the expressions for  the energy density and the pressure
\begin{equation}
\rho=z^{-0.5}\left[C+\mu^2 z\right]^{0.5} =\left[Ca^{-6}+\mu^2\right]^{0.5}
\end{equation}
and  
\begin{equation}
p=-\frac{\mu^2 z^{2\nu +0.5}}{\left(C+\frac{\mu^2}{2\nu+1}z^{2\nu+1}\right)^{0.5}}=-\frac{\mu^2 }{\left(Ca^{-6}+\mu^2\right)^{0.5}}, 
\end{equation}
respectively. The EoS parameter is given by
\begin{equation}
\omega=-\frac{\mu^2 z^{2\nu +1}}{C+\frac{\mu^2}{2\nu+1}z^{2\nu+1}}. 
\end{equation}
So this example corresponds to the usual Chaplygin gas \cite{Kamenshchik}. As is well known, in this case,  for small $a$ ($a^6<<C\mu^{-0.5}$), the energy density and the pressure take the forms, approximately,
\begin{equation}
\rho\approx C^{0.5}a^{-3},\quad p\approx 0 
\end{equation}
with $\omega=0$, which corresponds to a matter-dominated universe. For a large value $a$, it follows that\begin{equation}
\rho\approx \mu,\quad p\approx -\mu 
\end{equation}
that is, $\omega=-1$, which corresponds to a dark energy-dominated universe.   So this simple and elegant model smoothly interpolates between a dust dominated phase, where $\rho\approx C^{0.5}a^{-3}$, and a de Sitter phase, where $p\approx -\rho$, through an intermediate regime described by the EoS for stiff matter, $p=\rho$.

ii) Now let us consider the case when $U=\mu z^\nu$, where $\mu$ and $\nu$ are some real constants. In this case, Eq.(5.7) gives the following expression for  the energy density:
\begin{equation}
\rho=z^{-0.5}\left[C+\frac{\mu^2}{2\nu+1}z^{2\nu+1}\right]^{0.5} =\left[Ca^{-6}+\frac{\mu^2}{2\nu+1}a^{12\nu}\right]^{0.5}.
\end{equation}
Here $\nu$ must be negative, because otherwise, $a\rightarrow \infty$ implies $\rho\rightarrow \infty$, which is not the case for expanding Universe. The equation  (5.6) gives the expression for the pressure, 
\begin{equation}
p=-\frac{\mu^2 z^{2\nu +0.5}}{\left(C+\frac{\mu^2}{2\nu+1}z^{2\nu+1}\right)^{0.5}}=-\frac{\mu^2 a^{12\nu }}{\left(Ca^{-6}+\frac{\mu^2}{2\nu+1}a^{12\nu}\right)^{0.5}}. 
\end{equation}
The EoS parameter is given by
\begin{equation}
\omega=-\frac{\mu^2 z^{2\nu +1}}{C+\frac{\mu^2}{2\nu+1}z^{2\nu+1}}=-\frac{\mu^2 a^{12\nu }}{Ca^{-6}+\frac{\mu^2}{2\nu+1}a^{12\nu}} =-\frac{\mu^2 }{Ca^{-6(1+2\nu)}+\frac{\mu^2}{2\nu+1}}.
\end{equation}
The deceleration parameter $q$ has the expression
\begin{equation}
q=-\frac{\ddot{a}}{aH^2}=\frac{\rho+3p}{2\rho}=\frac{Ca^{-6}+\frac{2\mu^2(3\nu+2)}{2\nu+1}a^{12\nu}}{2(Ca^{-6}+\frac{\mu^2}{2\nu+1}a^{12\nu})}.
\end{equation}
For accelerating universe, $q$ must be negative i.e.,  $\ddot{a}>0$. Hence we have
\begin{equation}
a^{-6(2\nu+1)}<-\frac{2\mu^2(3\nu+2)}{C(2\nu+1)}.
\end{equation}
This means that for a small value of the scale factor we have a decelerating universe while for large values of scale factor we have an accelerating universe. The transition between these two phases occurs when the scale factor is equal to
\begin{equation}
a=a_c=-\left[\frac{2\mu^2(3\nu+2)}{C(2\nu+1)}\right]^{\frac{1}{-6(2\nu+1)}}.
\end{equation}

\section{Unified scalar--fermionic generalized Chaplygin gas model of dark energy and dark matter from g-essence with Yukawa interactions}

   In the previous section we constructed the unified scalar--fermionic Chaplygin gas model of dark energy and dark matter using the particular g-essence with  Yukawa interactions  having the scalar--fermionic DBI Lagrangian form (5.2). In this section we extend these results of the previous section for the scalar--fermionic generalized Chaplygin gas case. For this purpose, we consider  the particular g-essence model with Yukawa interactions which has the following scalar--fermionic DBI Lagrangian form
		\begin{equation}
K=U(1+V_1 X^{\frac{1}{2n}}+ V_2Y^{\frac{1}{n}})^n,
\end{equation}
where  		\begin{equation}
U=\eta \phi u, \quad V_1=V_1(\phi,\bar{\psi},\psi),\quad V_2=V_2(\phi,\bar{\psi},\psi).
\end{equation} Substituting this expression into (2.16)  we get 
	\begin{eqnarray}
p&=&U(1+V_1 X^{\frac{1}{2n}}+ V_2Y^{\frac{1}{n}})^n,\\ 
\rho&=&-U(1+V_1 X^{\frac{1}{2n}}+ V_2Y^{\frac{1}{n}})^{n-1}.	\end{eqnarray}
  These equations give
  	\begin{equation}
p=-\frac{(-U)^{\frac{1}{1-n}}}{\rho^{\frac{n}{1-n}}}=-\frac{(-U)^{\alpha+1}}{\rho^{\alpha}}=-\frac{A}{\rho^{\alpha}},
\end{equation}
 where $\alpha=n(1-n)^{-1}, \quad A=(-U)^{\alpha+1}$.  It is the scalar--fermionic generalized Chaplygin gas model \cite{Kamenshchik}. From (2.16) and (6.5) we get
   	\begin{equation}
\rho=a^{-3}\left[3(1+\alpha)\int (-U)^{\alpha+1}a^{3(1+\alpha)-1}da+C\right]^{\frac{1}{1+\alpha}}=z^{-\frac{1}{1+\alpha}}\left[C+\int (-U)^{\alpha+1}dz\right]^{\frac{1}{1+\alpha}}, 
\end{equation} 
where $C=const, \quad z=a^{3(1+\alpha)}$.	At the same time, for the pressure we obtain the following expression 
	\begin{equation}
p=-(-U)^{\alpha+1}z^{\frac{\alpha}{1+\alpha}}\left[C+\int (-U)^{\alpha+1}dz\right]^{-\frac{\alpha}{1+\alpha}}. 
\end{equation} 
These formulas give  the following expression for the EoS parameter:
	\begin{equation}
\omega=-\frac{ z(-U)^{\alpha+1}}{C+\int (-U)^{\alpha+1}dz}.
\end{equation}
Now we can consider different types solutions of the g-essence   equations. For example, if for simplicity, we consider the case $\omega=const$ then from  (6.8) we obtain the following expression for the function $U$:
 	\begin{equation}
U=-A_0^{\frac{1}{1+\alpha}}a^{-3(1+\omega)},
\end{equation}
 where $A_0$ is an integration constant. 	Let us consider some examples: i) if $\omega=-1$ (the de Sitter case) then $U=-A_0^{\frac{1}{1+\alpha}}$; 	ii) if $\omega=0$ (the dust  case) then $U=-A_0^{\frac{1}{1+\alpha}}a^{-3}$; iii) if $\omega=1$ (the stiff matter case) then  $U=-A_0^{\frac{1}{1+\alpha}}a^{-6}$ and so on. Now the construction of  solutions of the Friedmann, Klein--Gordon and Dirac equations is a formal problem so we omit it here.
 
	\section{Scalar--fermionic DBI generalization of the  g-essence model with Yukawa interaction (3.1)}
	Our aim in this section is to construct the scalar--fermionic DBI generalization of the  g-essence model with Yukawa interaction (3.1). To do it, we note that the model (3.1) is  some approximation of  the following scalar--fermionic DBI model:
	 \begin {equation}
K=\epsilon A\left\{\sqrt{1+A^{-1}[2\alpha_1 X+ 2\alpha_2 X^n+2\beta_1 Y+2\beta_2 Y^m]}-1\right\}+ V,
\end{equation}
where $\epsilon=+ 1, \quad A=const$ and
 \begin {equation}
V=\alpha_3 V_{1}(\phi)+\beta_3 V_2(\bar{\psi}, \psi)+\eta U_1(\phi)U_2(\bar{\psi}, \psi).
\end{equation} 
From (7.1)  we get
\begin{eqnarray}
	p&=&\epsilon A\left\{\sqrt{1+A^{-1}[2\alpha_1 X+ 2\alpha_2 X^n+2\beta_1 Y+2\beta_2 Y^m]}-1\right\}+ V,\\ 
	\rho&=&\frac{2\epsilon\alpha_2(n-1)X^n-\epsilon\beta_1 Y+\epsilon\beta_2(m-2)Y^m-\epsilon A}{\sqrt{1+A^{-1}[2\alpha_1 X+ 2\alpha_2 X^n+2\beta_1 Y+2\beta_2 Y^m]}}+\epsilon A-V.
	\end{eqnarray} 
	The study  the system of equations of g-essence (2.9)--(2.14) with the expressions for the pressure and the  energy density  given by (7.3)--(7.4) is a very hard job. Let us simplify the problem. Let $\alpha_2=\beta_1=0$ and $m=2$. Then equations (7.3)--(7.4) take the form
	\begin{eqnarray}
	p&=&\epsilon A\left\{\sqrt{1+A^{-1}[2\alpha_1 X+ 2\beta_2 Y^2]}-1\right\}+ V,\\ 
	\rho&=&-\frac{\epsilon A}{\sqrt{1+A^{-1}[2\alpha_1 X+ 2\beta_2 Y^2]}}+\epsilon A-V
	\end{eqnarray}
which corresponds to  the  EoS
	\begin{equation}
p=-\frac{A^2}{\rho-\epsilon A+V}-\epsilon A+V.\end{equation}The corresponding EoS parameter is given by
\begin{equation}
\omega=-\frac{A^2}{\rho(\rho-\epsilon A+V)}-\frac{\epsilon A-V}{\rho}
=\frac{A^2}{\epsilon A-V}\left(\frac{1}{\rho}-\frac{1}{\rho-\epsilon A+V}\right).\end{equation}
Now let us consider two particular limit cases when $\rho>>\epsilon A-V$ or $\rho<<\epsilon A-V$.

i) First we consider the case when 	 $\rho>>\epsilon A-V$.  Let us rewrite the equation (5.7)  as
	\begin{equation}
p=-\frac{A^2}{\rho}\left(\frac{1}{1-\frac{\epsilon A-V}{\rho}}\right)-\epsilon A+V.\end{equation}Hence,  in the first approximation,  we get
	\begin{equation}
p=-\frac{A^2}{\rho}-\epsilon A+V\end{equation}
which corresponds to the Chaplygin gas case \cite{Kamenshchik}. Now let us consider the second approximation. Then from (7.9) we obtain
\begin{equation}
p=-\frac{A^2}{\rho-\epsilon A+V}\approx -\frac{A^2}{\rho}\left(1+\frac{\epsilon A-V}{\rho}\right)-\epsilon A+V=-\frac{A^2}{\rho}-\frac{A^2(\epsilon A-V)}{\rho^2}-\epsilon A+V.\end{equation}

ii) Now let us  consider the second  case when 	 $\rho<<\epsilon A-V$.  Then from (7.7) we get
	\begin{equation}
p=-\frac{A^2}{\rho-\epsilon A+V}\approx \frac{A^2}{\epsilon A-V}\left(1+\frac{\rho}{\epsilon A-V}\right)-\epsilon A+V=\Lambda_g+\frac{A^2}{(\epsilon A-V)^2}\rho,\end{equation}
 where 
	\begin{equation}
\Lambda_g=\frac{A^2}{\epsilon A-V}-\epsilon A+V.\end{equation}
  In this case, the  EoS parameter is given by
\begin{equation}
\omega=\frac{A^2}{(\epsilon A-V)^2}+\frac{\Lambda_g}{\rho}.\end{equation}
  
  \section{Conclusion} 
	Recently, the so-called g-essence has been proposed to be a candidate of dark energy. It has the non-standard kinetic scalar and fermionic terms and has interesting properties. We expect that g-essence has rich properties and when it  is used in cosmology, its rich properties could have some interesting consequences. In this work, we found  the exact power law solution of the particular  g-essence model with the Yukawa scalar--fermionic interactions. The corresponding scalar and fermionic potentials are presented.  Of course, there might be other exact solutions of the g-essence different from the power law solution.  Anyway, our results obtained in the present work showed that the g-essence with the Yukawa interactions can describes the accelerated expansions of the universe. In fact, we have shown that some particular g-essence models with Yukawa  interactions correspond to the usual and generalized Chaplygin gas unified models of dark energy and dark matter.  Also we presented some scalar--fermionic DBI models which again can describe the unified dark energy--dark matter system.
	
	\section{Acknowledgement}
	We would like to thank the anonymous referees for providing us with constructive comments and suggestions to improve this work.

\end{document}